\documentclass[twocolumn]{aastex631}

\usepackage{float}

\def\etal{{et~al.\null}}

\graphicspath{{./}{figures/}}

\newcommand{\Gaia}{{\it Gaia}}

\newcommand{\kms}{{\>\rm km\>s^{-1}}}

\newcommand{\uBVI}{{\it uBVI}} 

\newcommand{\oCen}{$\omega$~Cen}
  
\shorttitle{Yellow PAGB Stars as Standard Candles}

\shortauthors{Ciardullo \etal}

\submitjournal{ApJ}
\accepted{April 12, 2022}
  
\begin{document}

\title{Yellow Post-Asymptotic-Giant-Branch Stars as Standard Candles. I\null. Calibration of the Luminosity Function in Galactic Globular Clusters}

\correspondingauthor{Robin Ciardullo}
\email{rbc@astro.psu.edu}
\author[0000-0002-1328-0211]{Robin Ciardullo}
\affil{Department of Astronomy \& Astrophysics, The Pennsylvania
State University, University Park, PA 16802, USA}
\affil{Institute for Gravitation and the Cosmos, The Pennsylvania
State University, University Park, PA 16802, USA}

\author[0000-0003-1377-7145]{Howard E. Bond}
\affil{Department of Astronomy \& Astrophysics, The Pennsylvania
State University, University Park, PA 16802, USA}
\affil{Space Telescope Science Institute, 3700 San Martin Dr., Baltimore, MD 21218, USA}

\author[0000-0002-8994-6489]{Brian D. Davis}
\affil{Department of Astronomy \& Astrophysics,
The Pennsylvania State University, 
University Park, PA 16802, USA}

\author[0000-0003-1817-3009]{Michael H. Siegel}
\affil{Department of Astronomy \& Astrophysics,
The Pennsylvania State University, 
University Park, PA 16802, USA}

\begin{abstract}

We use results of a survey for low-surface-gravity stars in Galactic (and LMC) globular clusters to show that ``yellow''  post-asymptotic-giant-branch (yPAGB) stars are likely to be excellent extragalactic standard candles, capable of producing distances to early-type galaxies that are accurate to a few percent.  We show that the mean bolometric magnitude of the 10 known yPAGB stars in globular clusters is $\langle M_{\rm bol} \rangle = -3.38 \pm 0.03$, a value that is $\sim$0.2~mag brighter than that predicted from the latest post-horizontal-branch evolutionary tracks. More importantly, we show that the observed dispersion in the distribution is only 0.10~mag, i.e., smaller than the scatter for individual Cepheids.  We describe the physics that can produce such a small dispersion, and show that, if one restricts surveys to the color range $0.0 \lesssim (B-V)_0 \lesssim 0.5$, then samples of non-variable yPAGB stars can be identified quite easily with a minimum of contamination.  The bright absolute $V$ magnitudes of these stars ($\langle M_V \rangle = -3.37$) make them, by far, the visually brightest objects in old stellar populations and ideal Population~II standard candles for measurements out to $\sim$10\,Mpc with current instrumentation.  A \textit{Hubble Space Telescope\/} survey in the halos of galaxies in the M81 and Sculptor groups could therefore serve as an effective cross-check on both the Cepheid and TRGB distance scales.

\end{abstract}

 \keywords{stars: AGB and post-AGB --- globular clusters: general --- galaxies: distances and redshifts}

\section{Introduction}
\label{sec:intro}

Our understanding of cosmology has come a long way in the past quarter century.  In the mid-1980's, the cosmic distance scale was uncertain to a factor of two, and astronomers struggled to reconcile measurements of the Hubble constant, the ages of globular clusters (GCs), and the matter density of the universe \citep[e.g.,][]{vandenberg90, Huchra92}.  Now, $H_0$ is known to much better than 10\% \citep[e.g.,][]{Freedman+01, Bennett+13, Riess+16, Planck+18}, dark matter has been modeled with precision \citep[e.g.,][]{Springel+05, Klypin+11, Vogelsberger+14}, and the problem of GC ages has been resolved through the discovery of dark energy \citep[e.g.,][]{Schmidt+98, Perlmutter+99}.

Nevertheless, there are inconsistencies, and the most disconcerting arguably involve our knowledge of the extragalactic distance scale and the expansion of the universe.  The physical scale of the baryonic acoustic oscillations, as measured by the \textit{WMAP\/} and \textit{Planck\/} spacecraft, is determined by simple physics and the expansion rate (i.e., the Hubble parameter) of the early universe.  If dark energy is the manifestation of a cosmological constant, the early-universe Hubble parameter can be tied directly to the present-day value of $H_0$; this relation, plus the curvature of the universe, implies a local value of $H_0 =67.37 \pm 0.54$~km\,s$^{-1}$\,Mpc$^{-1}$ \citep{Planck+18}.  However, determinations of $H_0$ based on Type~Ia supernovae (SNe~Ia), as calibrated by the Cepheid period-luminosity relation, yield significantly larger values, with two recent measurements giving $H_0 = 73.2 \pm 1.3$ and $73.04 \pm 1.04\rm\,\kms\,Mpc^{-1}$ \citep{Riess+21a, Riess+21b}.  Similarly, the tip-of-the-red-giant-branch (TRGB) calibration of SNe~Ia luminosities yields only marginal agreement with the microwave background, $H_0=69.6 \pm 0.8\,\rm (statistical) \pm 1.9\, (systematic)\,\kms\,Mpc^{-1}$ \citep{Freedman+20}, while measurements from megamasers \citep{Pesce+20} and gravitationally lensed quasars \citep{Wong+20} also support these higher values.  Either there is a systematic problem with the local scale of the universe, or the cosmological-constant interpretation of dark energy is incorrect and ``new physics'' is needed to understand the expansion history of the universe \citep[see, e.g.,][]{diValentino2021}.

One way to improve upon the current situation is to find additional calibrators for the intrinsic brightnesses of SNe~Ia. At first glance, there is no shortage of candidates, as the distance ladders presented by \citet{Ciardullo12} and \citet{deGrijs13} list over a dozen different methods that might serve to cross-check measurements of the local Hubble expansion.  However, to be competitive with the Cepheid and TRGB techniques, a standard candle must ideally (1)~have a strong theoretical basis, (2)~be well-calibrated via measurements within the Milky Way and Large Magellanic Cloud, (3)~be capable of reaching the nearest well-mixed clusters for the calibration of the surface-brightness-fluctuation method \citep{Blakeslee+21}, as well as Type~Ia supernovae, and (4)~have a small ($\lesssim$10\%) dispersion in their predicted luminosities.   These conditions present problems for most of the well-known distance methods.

Here we present an exploratory calibration of an extragalactic standard candle that shows promise for satisfying these criteria: the intermediate-temperature (yellow) post-asymptotic-giant-branch (hereafter yPAGB) stars of old stellar populations. In \S\ref{sec:theory}, we review the reasons why these stars should make excellent Population~II distance indicators, and
in \S\ref{sec:testing}, we test our conjecture using the $V$-band magnitudes and colors of low-surface-gravity stars in Galactic GCs.  In \S\ref{sec:candles} and \S\ref{sec:gaia}, we describe the best way to use yPAGB stars as standard candles and show that the observed scatter in their $V$-band magnitudes is better than that for Cepheids at a fixed period. In~\S\ref{sec:frequency}, we discuss the effect that stellar population has on the number of yPAGB stars in a galaxy and argue that a population's far-UV minus $V$-band color is the best predictor for the success of a yPAGB survey.  Finally, in \S\ref{sec:discuss}, we compare the efficiency of yPAGB surveys to those for Cepheids and the TRGB and discuss the utility of this new technique for the calibration of SNe~Ia luminosities.

\section{Theoretical Post-Horizontal-Branch Evolution}
\label{sec:theory}

\begin{figure}
\centering
\includegraphics[width=0.473\textwidth]{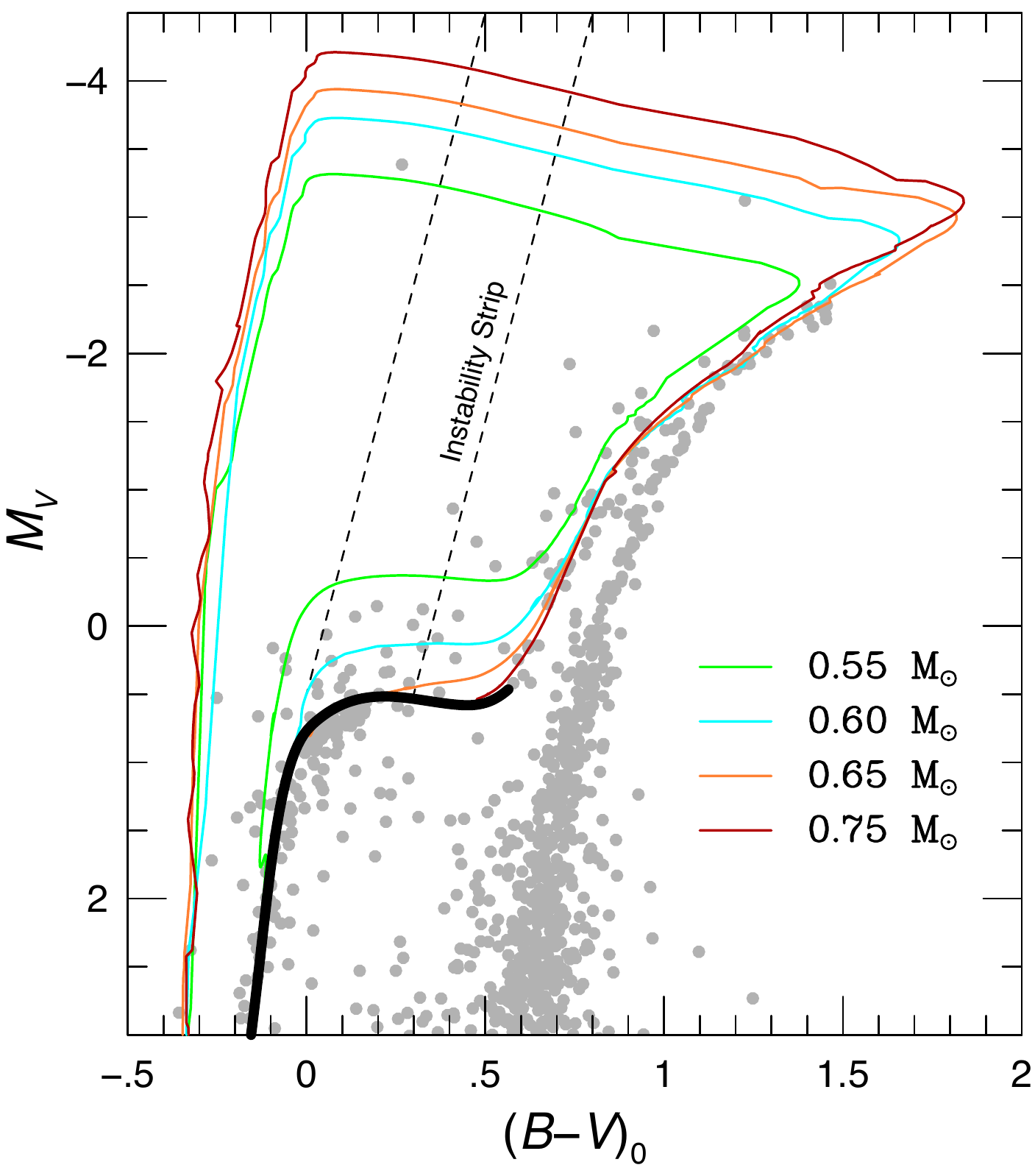}
\caption{Color-magnitude diagram showing four post-red-giant evolutionary tracks with metallicity $\rm[M/H] = -1.5$. The tracks are from \citet{Moehler+19}, and are for stars that arrive on the zero-age horizontal branch with masses of 0.55, 0.60, 0.65, and $0.75\,M_\odot$.  The thick black line traces the zero-age horizontal branch. Luminosity and effective temperature have been converted to visual absolute magnitude and $B-V$ color as described in the text. Underlying the tracks is the color-magnitude diagram of the globular cluster M79 \citep{Davis+22}.  The tracks illustrate how stars evolve off the zero-age horizontal branch to the base of the AGB, and then eventually to the post-AGB phase.  Note that M79 contains two stars that lie on post-AGB tracks. Post-AGB evolution for an individual stars is at a nearly constant bolometric luminosity, which depends on the star's core mass. The brightening in $M_V$ along the post-AGB tracks is due to the decreasing bolometric correction as the temperature increases. The approximate location of the Cepheid and RV~Tauri instability strip is indicated by the dashed lines.}
\label{fig:hrdiagram}
\end{figure}

The post-horizontal-branch (post-HB) evolution of low-mass stars is illustrated in the observational color-magnitude diagram (CMD) in Figure~\ref{fig:hrdiagram}. Here we plot the CMD (visual absolute magnitude, $M_V$, versus reddening-corrected $B-V$ color) for members of a typical GC, M79. The data are from \citet[][hereafter D22]{Davis+22}, and field interlopers have been removed as described in that paper. Superposed are theoretical post-HB evolutionary tracks\footnote{We thank Marcelo Miller Bertolami for sending us detailed tables of these tracks with a finer time resolution than given in the \citet{Moehler+19} paper.} from a recent paper by \citet{Moehler+19}. These tracks have been converted from luminosity and effective temperature to the observational quantities using the online PARSEC YBC web tool\footnote{\url{http://stev.oapd.inaf.it/YBC/}} \citep{Chen+19}.  For clarity, the tracks have been smoothed by omitting rapid excursions produced by helium shell flashes during late evolutionary stages, as discussed in more detail in D22.

Following the ignition of core helium burning at the tip of the RGB, a star moves to the zero-age horizontal branch (ZAHB), with its initial position determined by its envelope mass: objects with larger envelope masses reside on the red end of the ZAHB, while those with lower-mass envelopes inhabit the blue end of the sequence.  When core helium is exhausted, the star brightens due to the release of gravitational potential energy from the core and nuclear energy from the hydrogen-burning shell.  During this thick shell-burning phase, the star moves redward in the CMD to the base of the asymptotic giant branch (AGB), and then rapidly increases in luminosity.  Eventually the star's helium shell burning becomes unstable, and the envelope gets eaten away, both by mass being deposited onto the core and by material being lost in a superwind.  The process continues until the envelope mass is reduced to $\sim\! 0.005 \, M_{\odot}$, at which point the star begins its blueward traversal of the CMD and becomes a post-AGB star. Note that, as the figure shows, M79 contains two bright members that lie on post-AGB tracks, as discussed by \citet{Bond+16} and D22.

As Figure~\ref{fig:hrdiagram} illustrates, post-AGB stars reach their visually brightest magnitudes in the yPAGB stage, just to the blue of the Cepheid instability strip. The suggestion that these stars should have a narrow luminosity function, and thus could be useful as standard candles, was first laid out by \citet{Bond97a,Bond97b}, and we elaborate the arguments here.  The bolometric luminosity of a PAGB star on its final traversal of the CMD depends almost exclusively on its core mass \citep{Paczynski70, Vassiliadis+94, Blocker+95, MillerBertolami16}, and this core mass, in turn, is related to  the star's original main-sequence mass via the initial-final mass relation \citep[IFMR; e.g.,][]{Cummings+18, El-Badry+18}. As the existence of the second-parameter problem for Milky Way GCs demonstrates, there is a scatter in the final masses for a given initial mass: clusters that have similar ages and metallicities can have very different HB morphologies, and therefore different post-RGB masses \citep[e.g.,][]{Sandage+67, Torelli+19}.  Moreover, stars that undergo enhanced mass loss, due to binary interactions, rapid rotation, or simply stochastic processes during normal stellar evolution, may prematurely abort their AGB evolution, and create post-early asymptotic giant branch (PEAGB) objects with a wide range of luminosities.  However, unless the dispersion in the IFMR is so wide as to wipe out any systematic in the final white dwarf mass distribution, the overall relationship should hold: the older the stellar population, the lower the mean mass of its HB star envelopes, the lower the mean final mass of the population's stars, and the fainter the mean magnitude of these stars on their final traversal of the color-magnitude diagram.  Since the age of the universe is finite, this means that there should be a \textit{low-luminosity edge} to the distribution of PAGB luminosities which is defined by the IFMR of a galaxy's oldest stars.


At the same time, there are two selection effects which should impose a soft \textit{upper limit} to the luminosities of Population~II PAGB stars. Both effects stem from the strong inverse correlation between the luminosity of a PAGB star and its rate of evolution across the HR diagram \citep[see, for example,][]{Paczynski70, Vassiliadis+94, MillerBertolami16}.  The first is the observational bias against detecting rapidly evolving stars:  the more luminous the star, the less time it spends in the PAGB phase, and the lower the likelihood of its detection.   In a Population~II system, this correlation, coupled with the slow evolution of the main-sequence turnoff and the shallow slope of the low-mass end of the IFMR \citep[e.g.,][]{El-Badry+18, Cummings+18}, should produce an observed PAGB luminosity function that is biased toward low-luminosity objects.

Amplifying this effect is a related systematic associated with circumstellar extinction.  Low-mass PAGB stars, such as BD+39$^\circ$\,4926 \citep{Gezer+15}, BD+14$^\circ$\,3071 \citep{Bond20}, and the yPAGB star in M79 \citep{Bond+16} have been shown to have little to no circumstellar material. This is likely because their evolutionary timescale across the HR diagram is slow enough to allow dust produced on the AGB to disperse into space.  Conversely, in higher-luminosity, faster-evolving objects, the AGB ejecta are still in the vicinity of the star, even after the star has become hot enough to ionize its surroundings \citep[e.g.,][]{Herrmann+09, Davis+18}.  The dust associated with the circumstellar ejecta attenuates the stellar emission and again causes the luminosity distribution of observed PAGB stars to skew towards lower values. 

As a result of the combination of these effects, most of the PAGB stars of an old stellar populations should lie in a narrow band at the high-luminosity end of the HR diagram, some four magnitudes above the locus of HB stars.

\section{Observational Tests}
\label{sec:testing}

A straightforward test of the utility of yPAGB stars as standard candles is to examine the luminosity function of such stars in Galactic GCs.  This is now quite feasible: by combining \textit{Gaia} proper motions and parallaxes from Early Data Release~3 \citep[EDR3;][]{Gaia+21}, with a photometric technique for identifying luminous stars with low surface gravities \citep{Bond05}, it is relatively easy to identify PAGB star candidates and determine the likelihood of their cluster membership.  D22 did exactly this, using \Gaia\/ EDR3 astrometry, $uBVI$ photometry, and Gaussian-mixture modeling to identify a sample of 438 above-horizontal-branch (AHB) objects that are highly probable members of 104 Milky Way and LMC GCs. Among these stars, D22 identified 13 luminous objects as being PAGB star candidates.  

\begin{figure}
\centering
\includegraphics[width=0.473\textwidth]{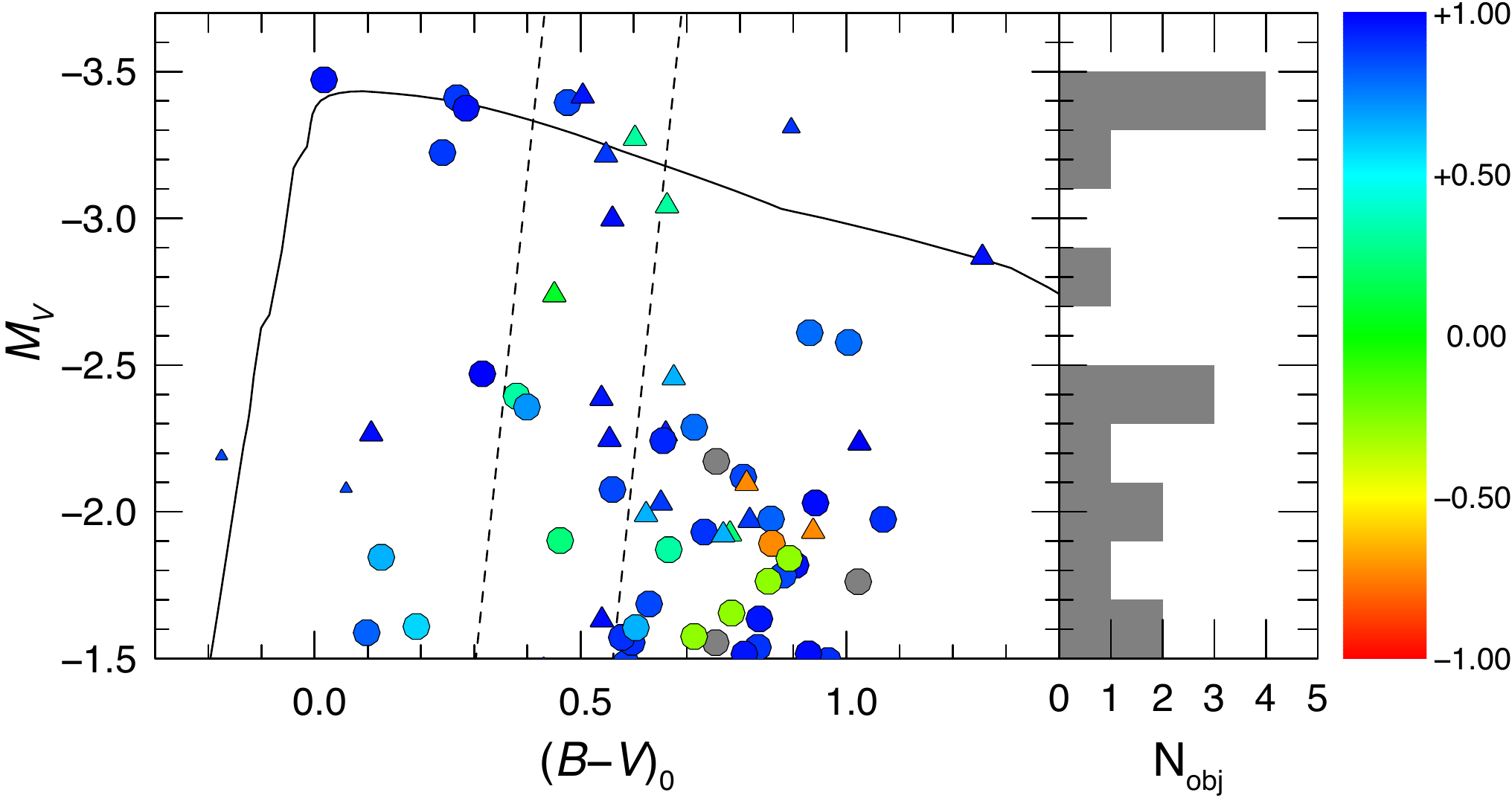}
\caption{Color-magnitude diagram of the brightest low-surface-gravity stars identified in the \citet{Davis+22} survey of Galactic and LMC globular clusters. Fill circles show non-variable stars and have typical errors (based on uncertainties in photometry, distance, and reddening) of $\sim 0.03$~mag in both $M_V$ and $(B-V)_0$. Large triangles represent mean $M_V$ values for RV Tauri and W Virginis variables and have slightly larger uncertainties, $\lesssim 0.05$~mag.  The small triangles denote single-phase measurements of variable stars.  The suspected RV~Tau variable with $M_V \simeq -3.3$ and $(B-V)_0 \simeq 0.9$ has a well-measured light curve in $V$ but no reliable $B$-band photometry.  The curve shows the expected location of stars with a constant bolometric luminosity of $\log L/L_{\odot} =  3.25$.  The approximate location of the instability strip is marked with dashed lines. Points are color-coded by the horizontal-branch morphology index of their host clusters (see text), as indicated by the color bar \citep{Lee+94, Borkova+00, Catelan09}; grey points identify stars in clusters without a well-measured HBR value.  The lack of PAGB stars from systems with red horizontal branches confirms the strong dependence of PAGB lifetime upon stellar mass.  The histogram on the right side of the figure represents the magnitude distribution for stars in the color range $0 \leq (B-V)_0 \leq 0.5$. }
\label{fig:pagb_cmd}
\end{figure}

To use these data, we first identified the known variable stars in the D22 sample, and replaced their single-epoch photometry with the stars' intensity-weighted mean magnitudes from \citet{Clement+01} and the references given in Table~\ref{tab:objects}.  We then took the photometric measurements of the stars, updated their estimated distances to the \textit{Gaia}+\textit{HST}+literature values given by \citet{Baumgardt+21}, and with two exceptions, applied the foreground extinction estimates summarized by 
\citet{Harris10}.\footnote{The extinction of M19 is spatially variable. Following \citet{Bond+21}, we estimated the reddenings of its individual PAGB and AHB stars using the high-resolution dust maps of  
\citet{Alonso-Garcia+12} and \citet{Johnson+17}.}
The result is the CMD shown in Figure~\ref{fig:pagb_cmd}.  Also plotted in the figure is the expected location of $\log L/L_{\odot} = 3.25$ stars (with $M = 0.56 \, M_{\odot}$ and [Fe/H] = $-1.5$) derived using the PARSEC web-tool for the calculation of bolometric corrections \citep{Chen+19}. 


Figure~\ref{fig:pagb_cmd} shows several interesting features.  The first is the color distribution of the PAGB stars.  Within the range $0.0 < (B-V)_0 < 1.0$, D22 identified 10 stars brighter than $M_V = -3.0$. Nine of these stars are within or blueward of the instability strip, and the one star redward of the instability strip is an RV Tau or semi-regular variable whose mean $(B-V)_0$ color is very poorly determined \citep{Clement+01, Wehlau+90, Jayasinghe+20}.  This distribution is as expected.  As detailed by \citet{Bond05}, the $uBVI$ photometric system is quite sensitive to surface gravity in the color range $0.0 < V-I < 0.9$, but the sensitivity to $\log g$ falls off rapidly for bluer objects, and tails off gradually towards the red.  For luminous stars, these $V-I$ limits are equivalent to the color range $0.0 < B-V < 0.7$ \citep{Worthey+11}, and this is precisely where the D22 post-AGB candidates are located.  Although the D22 sample also includes objects as red as $(B-V)_0 = 1.0$, only one of our post-AGB stars is located redward of the instability strip. 

Also, as expected, the magnitude separation between the PAGB stars and other AHB objects increases with effective temperature until $(B-V)_0 \simeq 0$.  This is a consequence of the evolutionary paths that stars take when leaving the ZAHB\null. Figure~\ref{fig:hrdiagram} illustrates that in the red, post-HB stars can attain luminosities that approach the AGB limit, making it difficult to distinguish between stars on their way to the AGB tip and post-AGB objects.  But blueward of the instability strip, the evolutionary paths of post-HB stars do not come within a magnitude of the PAGB locus (until the stars become so hot that the bulk of their emission is in the UV\null).  This suggests that the only possible contaminants in a PAGB survey (aside from foreground interlopers) are rapidly evolving stars undergoing thermal pulses near their surface, PEAGB stars evolved from objects at the extreme blue end of the HB, and various products of binary evolution.  As Figure~\ref{fig:pagb_cmd} demonstrates such objects must be extremely rare.  The D22 study included 104 of the oldest clusters in the Milky Way and LMC, and encompassed more than $\sim$$1.5 \times 10^7 L_{\odot}$ of $V$-band light.  Yet in the survey, there are no stars with colors between $0.0 < (B-V)_0 < 0.5$ that are within $\sim$0.7~mag of the PAGB track.

Perhaps the most interesting feature of Figure~\ref{fig:pagb_cmd} concerns its implications for the origins of PAGB stars and their rate of stellar evolution.  The GCs of the Milky Way and the LMC extend over $\sim$2.5~dex in metallicity \citep{Harris10}.  Yet as detailed in Table~\ref{tab:objects}, the metallicities of the clusters that contain PAGB stars brighter than $M_V = -3.0$ are relatively homogeneous, varying by less than a factor of $\sim$3. Moreover, these 10 PAGB stars belong to only seven GCs, as three of the systems possess two of these intermediate-temperature, low-surface-gravity objects. 

At first glance, these statistics seem peculiar, especially since, given a population of 10 A-, F-, or G-type PAGB stars, the binomial probability of 3 clusters out of 104 having two or more objects is less than 1\%.  However, the compilation by \citet{Harris10} shows that those eight clusters are among the most luminous GCs in the Milky Way, and contain $\sim$15\% of the GC system's total $V$-band light.  Similarly, almost $\sim$40\% of the total luminosity emitted by Milky Way GCs comes from clusters with $-1.29 \leq \textrm{[Fe/H]} \leq -1.74$.  So it should not be too surprising to find that PAGB stars are often found in these large, moderately metal-poor systems.

The most telling statistic, however, comes from the structure of the parent clusters' HBs.  We can quantify HB morphology using the index proposed by \citet{Lee+94}
\begin{equation}
    {\rm HBR} = \frac{B-R}{B+R+V} \, ,
\end{equation}
where $B$, $R$, and $V$ represent the numbers of HB stars blueward, redward, and inside the instability strip, respectively.   While imperfect \citep[see the discussion by][]{Torelli+19}, this index has been measured for most Milky Way and Magellanic Cloud GCs, allowing us to quickly identify clusters with predominantly red (HBR $< 0$) and blue (HBR $> 0$) HBs.  Based on the compilations of HBR values given by \citet{Lee+94}, \citet{Borkova+00}, and \citet{Catelan09}, slightly more than a third of the $V$-band luminosity of the Milky Way GC system comes from to clusters with red HBs.  Yet every one of the PAGB stars identified by D22 and in earlier studies belong to clusters with blue HBs.

The most natural way to explain this property is through the evolutionary timescale:  because red HB stars have more massive envelopes, they presumably create objects with higher PAGB masses, and therefore shorter PAGB evolutionary times.  The fact that we see no yPAGB stars in systems with red HBs is direct confirmation of the strong inverse relation between stellar mass and PAGB lifetime and/or the shallow slope of the low-mass end of the IFMR\null. It also supports our argument for a soft upper limit to the luminosities of Population~II PAGB stars.

Conversely, stars with small envelope masses produce relatively long-lived PAGB stars, and these objects occupy a distinct band in the color-magnitude diagram.  While extreme-HB stars \textit{could} evolve into objects that fill the void between the HB and PAGB regions of CMD \citep{Moehler+19}, such objects are not present in the D22 data set.  PEAGB stars must spend very little (if any) time in the ``yellow'' region of the HR diagram, either because of a rapid evolutionary timescale or an extreme dependence of post-HB luminosity on stellar mass.


\begin{deluxetable*}{llccccccl}
\tablewidth{0 pt}
\tabletypesize{\footnotesize}
\tablecaption{Yellow and Red PAGB Stars\label{tab:objects}}
\tablehead{
\colhead{Star\tablenotemark{\footnotesize a}} &
\colhead{$\alpha\rm(J2000)$} &
\colhead{$\delta\rm(J2000)$} &
\colhead{[Fe/H]\tablenotemark{\footnotesize b}}  &
\colhead{$E(B-V)$\tablenotemark{\footnotesize b}} &
\colhead{$(m-M)_0$\tablenotemark{\footnotesize c}} &\colhead{$(B-V)_0$} &
\colhead{$M_V$\tablenotemark{\footnotesize d}} &
\colhead{Photometry Reference}
}
\startdata
M79 PAGB  &05:24:10.36 &$-$24:29:20.7 &$-1.60$ &0.01  &15.58 &0.27  &$-3.41 \pm 0.04$ &\citet{Bond+16}\\
$\omega$ Cen V1 &13:26:05.17 &$-$47:23:42.5 &$-1.53$ &0.12 &13.67 &0.55 &$-3.21 \pm 0.04$ &\citet{Braga+20}\\
$\omega$ Cen HD 116745 &13:26:26.32 &$-$47:16:27.3 &$-1.53$ &0.12 &13.67 &0.24 &$-3.22 \pm 0.02$ &\citet{Davis+22}\\
M5 V84 &15:18:36.14 &+02:04:16.4 &$-1.29$ &0.03 &14.37 &0.66 &$-3.04 \pm 0.04$ &\citet{Rabidoux+10} \\
M5 V42 &15:18:24.78	&+02:02:53.2 &$-1.29$ &0.03 &14.37 &0.60 &$-3.27 \pm 0.03$ &\citet{Rabidoux+10} \\
NGC 5986 PAGB 1 &15:46:03.36 &$-$37:46:44.2 &$-1.59$ &0.29 &15.14  &0.02  &$-3.47 \pm 0.04$ &\citet{Davis+22} \\
NGC 5986 PAGB 2 &15:46:04.94 &$-$37:47:02.4 &$-1.59$ &0.29 &15.14 &0.28 &$-3.37 \pm 0.04$ &\citet{Davis+22} \\
M19 ZNG 4 &17:02:35.18 &$-$26:15:23.1 &$-1.74$ &0.34 &14.84 &0.48 &$-3.39 \pm 0.05$ &\citet{Bond+21}\\
M28 V17 &18:24:35.84 &$-$24:53:15.9 &$-1.32$ &0.40 &13.65 &0.90\rlap{\tablenotemark{\footnotesize e}}  &$-3.31 \pm 0.05$ &\citet{Jayasinghe+20}\\
M2 V11 &21:33:32.41	&$-$00:49:05.8 &$-1.65$ &0.06 &15.34 &0.50 &$-3.42 \pm 0.05$ &\citet{Demers69}\\
\enddata
\tablenotetext{a}{V: variable star from the list of \citet{Clement+01}; ZNG: \citet{Zinn+72} }
\tablenotetext{b}{Host cluster metallicity and reddening from  \citet{Harris10}, except the reddening for NGC\,5986 is from \citet{Alves+01}, and the reddening for M19 ZNG~4 is from \citet{Bond+21}  }
\tablenotetext{c}{Host cluster distance from \citet{Baumgardt+21}}
\tablenotetext{d}{The uncertainties on $M_V$ include the distance errors given by \citet{Baumgardt+21}, an estimate of the uncertainty in $E(B-V)$, and errors in the photometry.  For variable stars, the latter number   is that associated with the star's intensity-weighted mean magnitude.}
\tablenotetext{e}{The mean $(B-V)$ color of this object is poorly known}
\end{deluxetable*}


\section{Using PAGB Stars as Standard Candles}
\label{sec:candles}

When first formed, all ZAHB stars have an initial core mass of $\sim$$0.48\, M_{\odot}$ \citep[e.g.,][]{Sweigart+78}; the stars' final core masses depend of the masses of their envelopes. The more mass in the envelope, the more mass is accumulated onto the core during the AGB phase, and the brighter the star becomes.  Clusters with red HB stars should therefore produce very luminous (but short-lived) post-AGB stars, while systems with blue HBs should create relatively long-lived post-AGB objects at or near the low-luminosity limit for the class.

As Figure~\ref{fig:pagb_cmd} illustrates, the 10 PAGB stars identified by D22 are all from systems with very blue HBs and all have very similar bolometric luminosities.  If we ignore M28-V17, whose location on the CMD is poorly known due to a lack of coverage in the $B$-band, then the stars' scatter about the $\log L = 3.25$ line is just 0.04 in $\log L$ (or 0.10 in magnitude).  This consistency supports the hypothesis that PAGB stars are excellent standard candles for Population~II systems.  If the PAGB stars of other galaxies follow the same luminosity distribution as those in Milky Way GCs, then their distances can be measured to $\sim$2\% using just a handful of objects.

Of course, Figure~\ref{fig:pagb_cmd} also illustrates a major difficulty with the use of these stars:  five of the PAGB candidates shown in the figure are RV Tau variables.  These stars, which have amplitudes of more than 1~mag and periods ranging from one to two months \citep{Clement+01}\footnote{\url{http://www.astro.utoronto.ca/~cclement/cat/listngc.html}}, help define the PAGB evolutionary path across the HR diagram, since their intensity-weighted mean magnitudes are known from photometric monitoring (see the references in Table~\ref{tab:objects}).  However, because of their variability, the utility of these objects for extragalactic distance measurements is limited, as any determination of their mean magnitudes would require at least a dozen observations spaced over a minimum of three or four months.

The usefulness of red PAGB stars is also problematic, as the separation between true PAGB objects and stars still on their way to the AGB is relatively small.  In addition, as the detailed stellar-evolution sequences of \citet{Moehler+19} demonstrate, helium-shell thermal pulses on the AGB often produce small excursions in this part of the CMD that can further confuse the issue.  

On the other hand, the A- and F-type PAGB stars located blueward of the instability strip are ideal targets for study.  In addition to the very small scatter in their bolometric luminosities, these objects offer a number of practical advantages for extragalactic astronomy.

The first is that yPAGB stars are the visually brightest objects in Population~II systems, and easily stand out amidst their surroundings.  In bolometric terms, yPAGBs are just as luminous as stars near the tip of the AGB, but while AGB stars emit the bulk of the radiation in the red and infrared, yPAGB objects have their peak emission in the optical. As a result, in the $V$-band, yPAGB stars are a full magnitude brighter than the most luminous AGB objects, and $\sim$2~mag brighter than stars at the tip of the RGB\null.  At bluer wavelengths, the contrast between the PAGB and AGB is even greater \citep[see, e.g., the $u$-band image of M19 in Figure~2 of][]{Bond+21}.

A second advantage of yPAGB stars is that they are relatively unaffected by dust, both foreground and circumstellar. As Population~II objects, yPAGBs are generally not associated with regions of high interstellar reddening, and even in late-type galaxies, these objects should exist far away from places where internal extinction is important.   Moreover, as pointed out in \S\ref{sec:theory}, by the time a low-luminosity PAGB reaches a temperature of $T_{\rm eff} \simeq 6000$~K, virtually all of the mass lost by the star during the AGB should have dispersed into space.  Circumstellar extinction should therefore not affect our estimates of the objects' luminosities.

Perhaps most importantly, yPAGB stars with colors between $0.0 \lesssim (B-V)_0 \lesssim 0.5$ are easy to detect. As detailed above, these objects occupy a unique position in the HR diagram, roughly four magnitudes above the HB\null.  In addition, due to their extremely high luminosities and therefore low surface gravities, yPAGB stars have Balmer jumps that are up to $\sim$50\% larger that those of higher-gravity objects of the same effective temperature \citep{Bond05}.  As a result, it is relatively simple to identify yPAGB stars by comparing photometry through a filter located just shortward of the Balmer break to measurements longward of the feature.  Systematics due to possible reddening or metallicity effects can then be minimized through the use of a color-difference index, such as $c_2 = (u-B) - (B-V)$ (\citealt{Bond05}, D22).  

If we compute the mean $V$-band luminosity of the 10 stars in Table~1 using the bolometric corrections derived from the PARSEC database \citep{Chen+19}, then in the color range between $0.0 \leq (B-V)_0 \leq 0.5$, PAGB stars should have a mean absolute magnitude of $\langle M_V \rangle = -3.37 \pm 0.03$, where the uncertainty represents the standard deviation of the mean.  Alternatively, if we take an empirical approach and use the mean $V$-band magnitude of all the nonvariable GC stars with colors $0.0 \leq (B-V)_0 \leq 0.5$ and absolute magnitudes brighter than $M_V \simeq -2.7$ is then $\langle M_V \rangle = -3.37 \pm 0.05$.  Both of these estimates depend only on the distances to the GCs \citep[which, at present, are based on the \textit{Gaia}+\textit{HST}+literature estimates of][]{Baumgardt+21} and the foreground reddening values of \citet{Harris10}.  As the \textit{Gaia} distances to the clusters improve, so will our knowledge of the $V$ magnitudes of these stars.

We do note that over the color range $0.0 \lesssim (B-V)_0 \lesssim 0.5$, the bolometric correction applicable to yPAGB stars changes by $\sim$0.1~mag.  Thus, to reduce the scatter in the stars' absolute $V$ magnitudes, one could apply a color-dependent bolometric correction to each star.  However, while such corrections are straightforward, they are likely not necessary.  Just as the color dependence of Cepheid periods is rendered moot by averaging over an ensemble of objects distributed throughout the instability strip \citep{Jacoby+92}, so is the color dependence of yPAGB magnitudes.  As long as the sample of objects numbers more than $\sim$10, this additional step would not significantly improve the accuracy of the technique.

\section{Confirmation from {\em Gaia\/} EDR3 \label{sec:gaia} }

\begin{figure}
\centering
\includegraphics[width=0.473\textwidth]{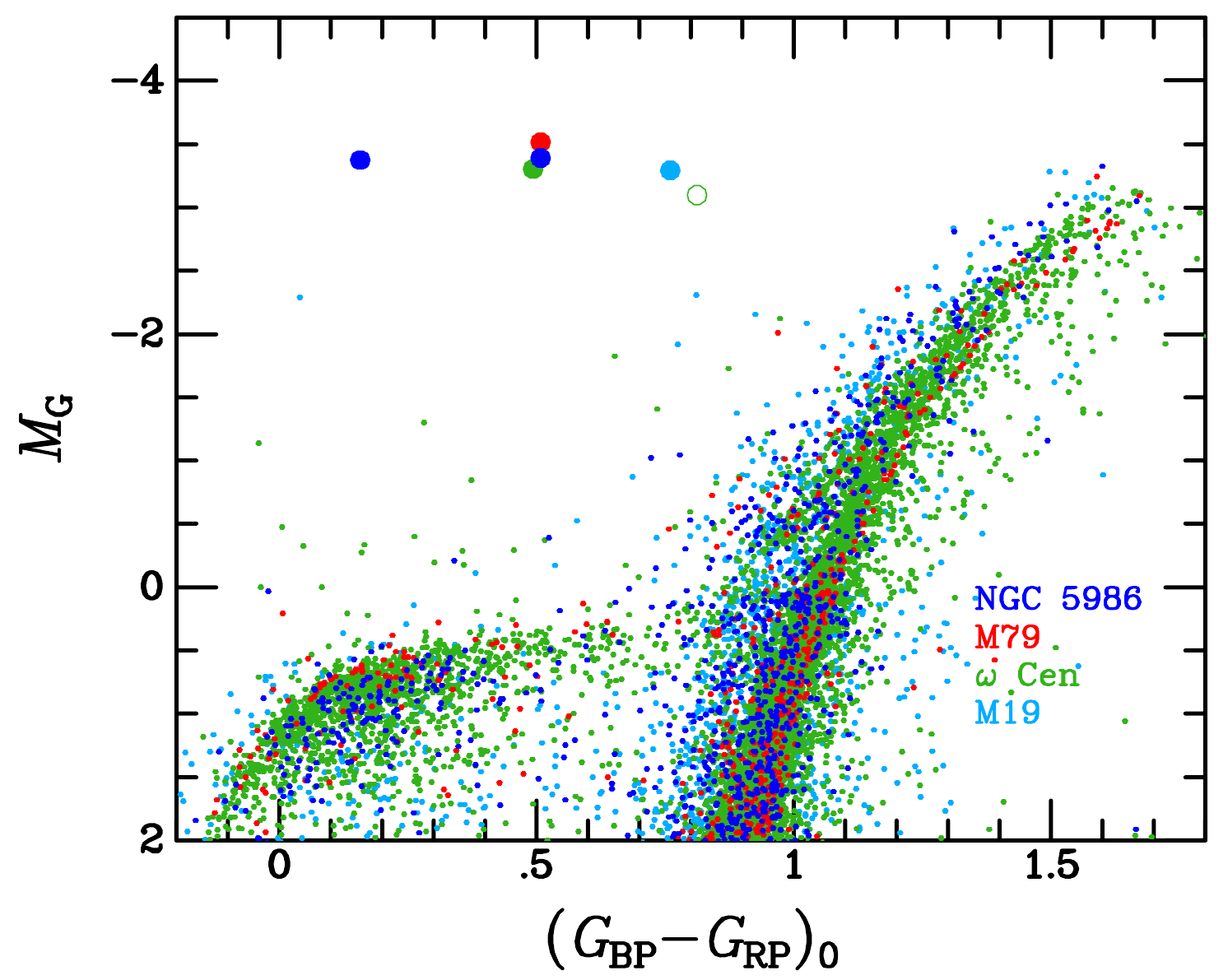}
\caption{Color-magnitude diagram for the combined membership of the four Milky Way globular clusters that contain nonvariable yellow PAGB stars.  The data are derived from \textit{Gaia\/} EDR3 photometry, and are color-coded as indicated in the legend. Field stars have been removed from the diagram based on proper motions and parallaxes, and the magnitudes and colors are corrected for distance and reddening, as described in the text. Filled circles mark the five yellow PAGB stars, color-coded according to cluster membership; the open circle is the RV Tau variable V1 in $\omega$~Cen. The figure illustrates the extremely narrow range of absolute magnitudes for the nonvariable yPAGB objects, and the large gap between them and the next brightest stars in the clusters.}
\label{fig:gaia_cmd}
\end{figure}

To further illustrate the potential of yPAGB stars as standard candles, we can use photometric data from 
\Gaia\/ EDR3\footnote{We obtained the \Gaia\/ EDR3 data from the Strasbourg Astronomical Data Center, at \url{https://vizier.cds.unistra.fr/viz-bin/VizieR}} for the four GCs that contain nonvariable yPAGB objects. 
Figure~\ref{fig:gaia_cmd} displays the combined \Gaia\/ CMD for M79, $\omega$~Cen, NGC\,5986, and M19, plotting absolute $G$ magnitude versus dereddened $G_{\rm BP}-G_{\rm RP}$ color. To select cluster members for this figure, we chose all stars lying within $6'$ of the cluster centers ($15'$ for \oCen), and eliminated most field interlopers by requiring the proper motions to lie within $\pm$$1.5\rm\,mas\,yr^{-1}$ of the mean cluster value in both coordinates ($\pm$$2.5\rm\,mas\,yr^{-1}$ for \oCen), and have parallaxes within $\pm$0.5~mas of the cluster mean.

To correct the data for extinction, we adopted the $E(B-V)$ values listed in Table~\ref{tab:objects}. Based on calculations given in the PARSEC online CMD tool\footnote{\url{http://stev.oapd.inaf.it/cmd}} \citep{Pastorelli2019}, we assumed extinction coefficients of $A_G=2.5924\,E(B-V)$ and $E(BP-RP) = 1.3918\, E(B-V)$.\footnote{These coefficients were calculated for the SED of a G2\,V star. They are reasonable for the yPAGB stars, but will have systematic offsets for very blue and very red objects.} As in Table~\ref{tab:objects}, the absolute distances for the clusters were taken from the recent compilation\footnote{Available online at \url{https://people.smp.uq.edu.au/HolgerBaumgardt/globular/}} of \citet{Baumgardt+21}.

In the figure, we plot the yPAGB stars in these four clusters as large filled circles; also included in the figure as an open circle is the RV~Tau variable V1 in \oCen. The figure supports the discussion in the previous two sections. Specifically, (1)~the five nonvariable yPAGB stars have a very narrow luminosity function\footnote{Note, in particular, that the two yPAGB stars in NGC\,5986 (which are at the same distance and suffer the same extinction) have $G$ magnitudes that differ by only 0.016~mag.} (rms scatter 0.09~mag; mean absolute magnitude $M_G=-3.37\pm 0.04$); and (2)~there is a gap of more than 1~mag between these objects and the nearest AHB stars.  Since the blue end of the ZAHB is well populated in all four clusters, it is safe to conclude that these stars define the low-luminosity end of the yPAGB stars' luminosity function. 

\section{Specific Frequency of Yellow PAGB stars}
\label{sec:frequency}

When planning extragalactic surveys, it is helpful to know how many targets one might reasonably expect to find in a single observation.  Though this number is difficult to calculate in general, there are ways to estimate the likelihood of a successful yPAGB survey.

From the ``fuel-consumption theorem,'' the number of stars in 
any post-main sequence phase of evolution, $N$, is
\begin{equation}
N = L_{\rm pop} \, B(t) \, \tau 
\label{eq:fct}
\end{equation}
where $L_{\rm pop}$ is the bolometric luminosity of the stellar population being surveyed, $B(t)$ is the bolometric-luminosity-specific stellar-evolutionary flux, and $\tau$ is the mean lifetime of the evolutionary phase being investigated \citep{Renzini+86, Buzzoni+06}. For all populations older than $\sim$3~Gyr, $B \simeq 1.8 \times 10^{-11}$~stars~yr$^{-1}~L_{\odot}^{-1}$, and the applicable bolometric correction is roughly $-0.85$ \citep{Buzzoni+06}.  Thus, the production rate of PAGB stars per unit $V$-band luminosity can be estimated readily from the integrated brightness of a target population.  The total yPAGB star population then only depends on the lifetime of the stage.

Unfortunately, for a mixed stellar population, the lifetime of PAGB stars is ill-defined, as the evolutionary timescale for these objects is a steep function of their core mass.  For a population mix similar to that of the Milky Way GCs, we can estimate $\tau$ directly.  Based on the GC luminosities given by \citet{Harris10} and \citet{Pessev+08}, the D22 \uBVI\/ survey included $\sim$$1.5 \times 10^7 L_{\odot}$ of $V$-band GC light.  Since the survey detected five nonvariable yPAGB stars, Equation~(\ref{eq:fct}) implies that the mean lifetime of these objects is of order $\sim$9000~yr, with a $1\,\sigma$ confidence interval extending from 5100 to 15,000~yr \citep{Gehrels86}.  This is consistent with the value one would infer from the $M = 0.55 \, M_{\odot}$, [Fe/H] $ = -1.5$ evolutionary track of \citet{MillerBertolami16}, shown in Figure~\ref{fig:hrdiagram}. 

Of course, as discussed in \S\ref{sec:testing}, all of the PAGB stars found by D22  come from clusters with blue HBs; the higher-mass PAGB objects produced in red-HB clusters presumably have much shorter evolutionary times and are therefore less common. So the actual lifetime of the yPAGB stars plotted in Figure~\ref{fig:hrdiagram} must be greater than that estimated above.  Thus a proper estimation for the number of yPAGB stars expected in a galaxy must consider the mix of stellar populations and HB morphologies present in the target population.  Such data will not generally be available for extragalactic investigations.

However, there is one observable that should be a good predictor for the number of PAGB stars present in an early-type system: the stellar population's UV upturn.  Traditional stellar population models predict that Population~II systems should have very little flux in the far UV\null.  However, satellites such as \textit{OAO-2\/} \citep{Code+82} and \textit{IUE\/} \citep[e.g.,][]{Oke+81, Burstein+88} demonstrated that this is not the case:  the spectral-energy distributions of elliptical galaxies and spiral bulges often increase shortward of $\sim$2500\,\AA\ \citep[e.g.,][]{Greggio+99, OConnell99}.  The two most plausible sources of this light are extreme HB stars and hot PAGB objects \citep{OConnell99, Yi08}, and, as D22 and Figure~\ref{fig:pagb_cmd} demonstrate, the former population generates the latter.   Thus we should expect the number of yPAGB stars to correlate strongly with a system's UV upturn.  This means that in systems such as M32 and NGC\,147 which have very weak UV excesses and relatively few blue HB stars \citep{Brown+00, Geha+15}, we should expect yPAGB stars to be relatively rare.  Conversely, in a galaxy such as NGC\,185, which has a significant blue HB population \citep{Geha+15}, we should see many more yPAGB  objects per unit stellar luminosity.  We should also expect the number density of yPAGB stars in the halo of M31 to follow that of the blue HB stars measured by \citet{Williams+15}.


\section{Discussion}
\label{sec:discuss}

If the PAGB luminosity limit is really as constant as the data suggest, then yPAGB stars are competitive with the very best extragalactic standard candles.  Cepheid variables are generally considered to be the gold standard for calibrating the luminosities of SNe~Ia, but their observation is relatively expensive, as at least a dozen visits are required to accurately determine their periods.  Moreover, optical observations of a single Cepheid variable can only yield a distance accurate to $\sim$10\% \citep{Riess+19}; it is only by measuring large numbers of Cepheids in a single galaxy that this uncertainty can be beaten down.  Thus, the technique works best in large late-type systems with high star-formation rates, large numbers of Cepheid variables---and, unfortunately, significant interstellar extinction. 

Alternatively, the TRGB method can produce an accurate distance to a Population~II system in a single visit. However, care must be taken to avoid the systematic effects introduced by image crowding \citep{Freedman21}, and the fading of the $I$-band RGB tip with metallicity \citep[e.g.,][]{Beaton+18}.  The technique is therefore best suited for small galaxies with exclusively metal-poor populations. 

In contrast, crowding should not be a major issue for yPAGB stars, since in the $V$-band, these objects are $\sim$2~mag brighter than the TRGB and (due to their smaller bolometric correction) one magnitude brighter than the most luminous AGB stars.  Moreover, we expect that the low-luminosity locus of yPAGB stars will be rather insensitive to galactic metallicity.  Based on the data in hand, we cannot say for certain how much the PAGB luminosity limit will change with metal abundance, as the sample of GCs hosting PAGB stars only spans the range $-1.74 \lesssim \textrm{[Fe/H]} \lesssim -1.29$ (Table~\ref{tab:objects}).    However, since the luminosity limit is produced by a galaxy's oldest stars, it is likely that the range of metallicity for the lowest-luminosity yPAGB objects in other galaxies will also be limited.  So we may not see much of a change in $\langle M_V \rangle$, even in environments such as the halos of large spiral galaxies.

The efficient detection of yPAGB stars in other galaxies does require observations through a $u$-band filter whose bandpass is almost entirely blueward of the Balmer break.  The F336W filter of the WFC3/UVIS camera of the \textit{Hubble Space Telescope\/} satisfies this criterion, and the instrument is efficient enough to reach the yPAGB stars of the M81 and Sculptor groups in as little as four spacecraft orbits.  Thus the method can provide an independent cross-check of the TRGB method in systems that are inaccessible to Cepheids.  Moreover, with longer exposures, WFC3 F336W detections can be extended outward to $\sim$10\,Mpc, enabling the direct calibration of several SNe~Ia. 

Our understanding of the luminosities of yPAGB stars is still in its earliest stages.  To confirm their utility as standard candles, additional observations in the various populations of the Local Group are needed.  But based on the results from the Milky Way's GCs, such additional studies are warranted, as the objects may provide a simple and direct link between direct trigonometric parallaxes from \textit{Gaia\/}  and the luminosities of SNe~Ia.



\begin{acknowledgments}

This research has made use of the VizieR catalogue access tool, CDS,
Strasbourg, France (DOI : 10.26093/cds/vizier). The original description 
of the VizieR service was published in 2000, A\&AS 143, 23.
 
 This work has made use of data from the European Space Agency (ESA) mission
{\it Gaia\/} (\url{https://www.cosmos.esa.int/gaia}), processed by the {\it Gaia\/}
Data Processing and Analysis Consortium (DPAC,
\url{https://www.cosmos.esa.int/web/gaia/dpac/consortium}). Funding for the DPAC
has been provided by national institutions, in particular the institutions
participating in the {\it Gaia\/} Multilateral Agreement.

The Institute for Gravitation and the Cosmos is supported by the Eberly College of Science and the Office of the Senior Vice President for Research at the Pennsylvania State University.  This research has made use of the NASA/IPAC Infrared Science Archive, which is funded by the National Aeronautics and Space Administration and operated by the California Institute of Technology.

\end{acknowledgments}


\bibliography{pagb.bib}

\end{document}